\documentclass[3p,times]{elsarticle}

\usepackage{ecrc}
\usepackage{amssymb}
\usepackage{amsmath}
\usepackage{lineno}
\usepackage{slashed}
\usepackage{graphicx}
\usepackage{atlasphysics}
\usepackage[bookmarks=false]{hyperref}
\usepackage{multirow}
\newcommand*\patchAmsMathEnvironmentForLineno[1]{%
  \expandafter\let\csname old#1\expandafter\endcsname\csname #1\endcsname
  \expandafter\let\csname oldend#1\expandafter\endcsname\csname end#1\endcsname
  \renewenvironment{#1}%
     {\linenomath\csname old#1\endcsname}%
     {\csname oldend#1\endcsname\endlinenomath}}% 
\newcommand*\patchBothAmsMathEnvironmentsForLineno[1]{%
  \patchAmsMathEnvironmentForLineno{#1}%
  \patchAmsMathEnvironmentForLineno{#1*}}%
\AtBeginDocument{%
\patchBothAmsMathEnvironmentsForLineno{equation}%
\patchBothAmsMathEnvironmentsForLineno{align}%
\patchBothAmsMathEnvironmentsForLineno{flalign}%
\patchBothAmsMathEnvironmentsForLineno{alignat}%
\patchBothAmsMathEnvironmentsForLineno{gather}%
\patchBothAmsMathEnvironmentsForLineno{multline}%
}
\newcommand{\sqrts}{\mbox{$\sqrt{s}$}}
\newcommand{\sqrtsnn}{\mbox{$\sqrt{s_{NN}}$}}
\newcommand{\Rcollcent}{\mbox{$R_{\mathrm{coll}}$}}

\newcommand{\Npart}{\mbox{$N_{\mathrm{part}}$}}
\newcommand{\ANpart}{\mbox{$\langle N_{\mathrm{part}}\rangle$}}
\newcommand{\Ncoll}{\mbox{$N_{\mathrm{coll}}$}}

\newcommand{\AuAu}{\mbox{Au+Au}}
\newcommand{\PbPb}{\mbox{Pb+Pb}}
\newcommand{\pp}{\mbox{\emph{p}+\emph{p}}}

\newcommand{\Rcp}{\mbox{$R_{\rm CP}$}}

\newcommand{\ETfcal}{\mbox{$\Sigma E_{\mathrm{T}}^{\mathrm{FCal}}$}}

\newcommand{\pID}{\mbox{$p_{\mathrm{ID}}$}}

\newcommand{\Deltaploss}{\mbox{$\Delta p_{\mathrm{loss}}$}}

\newcommand{\Comp}{\mbox{$C$}}
\newcommand{\avgC}{\mbox{$\langle C\rangle$}}

\newcommand{\fs}{\mbox{${f_{\mathrm{S}}}$}}

\newcommand{\Ktopi}{\mbox{$K/\pi$}}

\volume{00}

\firstpage{1}

\journalname{Nuclear Physics A}

\runauth{ATLAS Collaboration}

\jid{nupha}

\usepackage{amssymb}
\usepackage[figuresright]{rotating}

\begin{document}

\begin{frontmatter}



\dochead{}

\title{Open heavy flavour production via semi-leptonic decay muons in
\PbPb~collisions at $\sqrtsnn = 2.76$~\TeV\ with the ATLAS detector at the LHC}


\author{Yujiao Chen on behalf of the ATLAS Collaboration}

\address{yujiao@phys.columbia.edu, Columbia University}

\begin{abstract}
Measurements of heavy quark production and suppression in
ultra-relativistic nuclear collisions probe the interactions of
heavy quarks with the hot, dense medium created in the
collisions. ATLAS has measured heavy
quark production in $\sqrtsnn = 2.76$~\TeV~ \PbPb~ collisions via
semi-leptonic decays of open heavy flavour hadrons to muons.
Results are presented for the per-event muon yield as
a function of muon transverse momentum, \pT, over the range of $4 < \pT
< 14$~\GeV. 
%Over that momentum range single muon production 
%results largely from heavy quark decays. 
The centrality dependence of the
muon yields is characterized by the ``central to
peripheral'' ratio, \Rcp. Using this measure, muon production
from heavy quark decays is found to be suppressed by a
centrality-dependent factor that increases smoothly from peripheral to
central collisions. Muon production is suppressed by approximately a
factor of two in central collisions relative to peripheral
collisions.  Within the experimental errors, 
the observed suppression is independent of muon \pT\ for all
centralities.
\end{abstract}


\end{frontmatter}

%%\linenumbers

%\section{ }
%\label{}
\section{Introduction}
Collisions between lead ions at the LHC are thought to produce
strongly interacting matter at temperatures well above the QCD
critical temperature. At such temperatures, strongly interacting
matter is expected to take the form of 
``quark-gluon plasma.'' High-\pT\ quarks and gluons generated in
hard-scattering processes during the initial stages of the nuclear
collisions are thought to lose energy in the quark-gluon plasma
resulting in ``jet quenching'' \cite{Majumder:2010qh} . Since the energy
loss results from the interaction of a quark or gluon with the
medium, jet quenching is thought to provide a valuable tool for
probing the properties of the quark-gluon plasma
\cite{Wang:1994fx,Baier:1998yf,Gyulassy:2000fs}. 
%% Measurements of 
%% heavy quarks are an important complement to studies of
%% light quark and gluon quenching. 
The contributions from radiative \cite{Djordjevic:2003zk}
and collisional \cite{Wicks:2005gt}
energy loss in weakly coupled calculations are expected to 
depend on the heavy quark mass.
In particular, the mass of heavy quarks is expected to reduce radiative
energy loss for quark transverse momenta less than or comparable to
the quark mass ($m$), $\pt \lesssim m$, through the dead-cone
effect \cite{Dokshitzer:2001zm}. However, measurements of heavy quark
production at RHIC via semi-leptonic decays to electrons showed a
combined charm and bottom suppression in \AuAu\ collisions comparable
to that observed for inclusive hadron production
\cite{Adare:2006nq,Abelev:2006db,Adare:2010de}. There is
disagreement in the theoretical literature regarding the
interpretation of the RHIC heavy quark suppression measurements
\cite{Djordjevic:2011tm,Gossiaux:2008jv,Uphoff:2011ad}
particularly regarding the role of non-perturbative effects
\cite{Horowitz:2008ig}. It is clear that
measurements of heavy quark quenching at the LHC are an essential
complement to inclusive jet  or
single hadron measurements. 
 
This paper presents results on heavy-quark production 
through muon semi-leptonic decay, using the ATLAS detector
~\cite{Aad:2008zzm} and approximately $7~{\rm \mu b^{-1}}$
of \PbPb\ data at $\sqrtsnn = 2.76$~\TeV\ collected in 
2010. Further details of the analysis can be found in
~\cite{ATLAS-CONF-2012-050}. 
The measurements were performed for several intervals of
collision centrality over the muon transverse momentum
range $4 < \pt < 14$~\GeV\ and the yields compared to those 
in a peripheral bin using \Rcp,
$\Rcp = \frac{\langle\Ncoll^{\mathrm{periph}}\rangle \mathrm{d}n^{\mathrm{cent}}}{\langle\Ncoll^{\mathrm{cent}}\rangle \mathrm{d}n^{\mathrm{periph}}}$, where $\mathrm{d}n^{\mathrm{cent}}$ and $\mathrm{d}n^{\mathrm{periph}}$ represent
the per-event differential rate for the same
observable in central and peripheral collisions,
respectively. $\langle\Ncoll^{\mathrm{cent}}\rangle$ and $\langle\Ncoll^{\mathrm{periph}}\rangle$
represent $\langle\Ncoll\rangle$ values calculated for the corresponding centralities.

\section{Experimental setup and muon reconstruction}
\label{sec:eventcent} 
%% The measurements presented in this paper were obtained using the
%% ATLAS muon spectrometer (MS), inner detector (ID), calorimeter, trigger and data
%% acquisition systems. 
%% A detailed description of these detectors and
%% their performance in proton-proton collisions can be found in 
%% Refs.~\cite{Aad:2008zzm,Aad:2009wy}. 
Muons were detected by
combining independent measurements of the muon trajectories from the
inner detector (ID) and the muon spectrometer (MS), which cover
$|\eta|<2.5$ and $|\eta|<2.7$ in pseudorapidity respectively. 
Two forward calorimeters placed symmetrically with respect to 
the collision point
and covering $3.2 < |\eta| < 4.9$ are used to
characterize \PbPb\ collision centrality. 
Minimum bias \PbPb\ collisions were identified using measurements from
the zero degree calorimeters (ZDCs) and the minimum-bias trigger
scintillator (MBTS) counters. Cuts are imposed 
both online and offline to reject background
and ensure selection of heavy ion collision events.
In total, 53 million minimum-bias \PbPb\ events are
selected.
Previous studies \cite{ATLAS:2011ah} indicate that the criteria 
used to select minimum-bias hadronic
\PbPb\ collisions have an efficiency of $98 \pm 2\%$. 
The centrality of \PbPb\ collisions was characterized by \ETfcal, the
total transverse energy measured in the forward calorimeters (FCal). 
%% For the
%% results presented in this paper, the minimum-bias
%% \ETfcal\  distribution was divided into centrality intervals according 
%% to the following percentiles of the \ETfcal\ distribution ordered from
%% the most central to the most peripheral collisions: 0-10\%,
%% 10-20\%, 20-40\%, 40-60\%, and 60-80\%. The centrality intervals were
%% determined after accounting for very peripheral events lost due to the
%% minimum-bias trigger efficiency.
A standard Glauber Monte-Carlo analysis was
used to estimate the average number of  participating
nucleons, $\langle \Npart \rangle$, and the average number of
nucleon-nucleon collisions, $\langle \Ncoll \rangle$, for
\PbPb\ collisions in each of the centrality bins. 
%%The results are shown in Table~\ref{tbl:glauber}. 
The \Rcp\ measurements presented
in the paper use the 60-80\% centrality bin as a common peripheral
reference. 

The performance of the ATLAS detector and offline analysis in
measuring muons in \PbPb\ collisions was evaluated using a Monte Carlo
(MC) data set 
\cite{atlassim}.  The MC data set was obtained by overlaying
GEANT4-simulated \cite{Agostinelli:2002hh} $\sqrts =
2.76$~\TeV\ \pp\ dijet events on to 1 million GEANT4-simulated
minimum-bias \PbPb\ events obtained from version 1.38b of the HIJING
event generator \cite{Wang:1991hta}. HIJING was run with default
parameters, except for the disabling of jet quenching. To simulate the
effects of elliptic flow  in \PbPb\ collisions, a parameterized 
$\cos{2\phi}$ modulation that varies with
centrality, $\eta$ and \pT\  was imposed on the
particles after generation.  
The efficiency for reconstructing muons associated with heavy flavour
decays was evaluated using the MC sample described above for different
bins in collision centrality. 

%% \begin{figure}
%% \centerline{
%% \includegraphics[width=\fighalfwidth]{figures/TemplateMCOverlayFigureForPaper}
%% }
%% \caption{
%% Simulated \Comp\ distributions, $\mathrm{d}P/\mathrm{d}\Comp$, for signal and background muons
%% for the full (0-80\%) centrality range and for the 0-10\%
%% and 60-80\% centrality bins for muons in the momentum
%% range of 5 -- 6 GeV.
%% }
%% \label{fig:composite}
%% \end{figure}
\section{Muon signal extraction and results}
The measured muons consist of both ``signal'' muons and background muons.
The signal muons are those that originate directly from the
\PbPb\ collision, from vector meson decays, and from heavy quark
decays. The background muons arise from pion and kaon decays, muons
produced in hadronic showers in the calorimeters, and mis-associations of
MS and ID tracks. 
%% Previous studies have shown that the signal and
%% background contributions to the reconstructed muon sample can be 
%% discriminated statistically
%% \cite{Aad:2011rr,ATLAS-CONF-2011-003}. 
For this analysis a weighted combination of two discriminant quantities has
been used  to from a ``composite'' discriminant, \Comp\ .
These two quantities are: a) 
the fractional momentum
imbalance, $\Deltaploss/\pID$, which quantifies the difference between the ID
and MS measurements of the muon momentum after accounting for the
energy loss of the muon in the calorimeters, and b)
the ``scattering significance'',
$S$, which characterizes deflections in the trajectory resulting from
(e.g.) decays in flight. Further details are given in~\cite{ATLAS-CONF-2012-050}.

\begin{figure}[t]
  \begin{tabular}{cc}
    \includegraphics[width=0.45\textwidth]{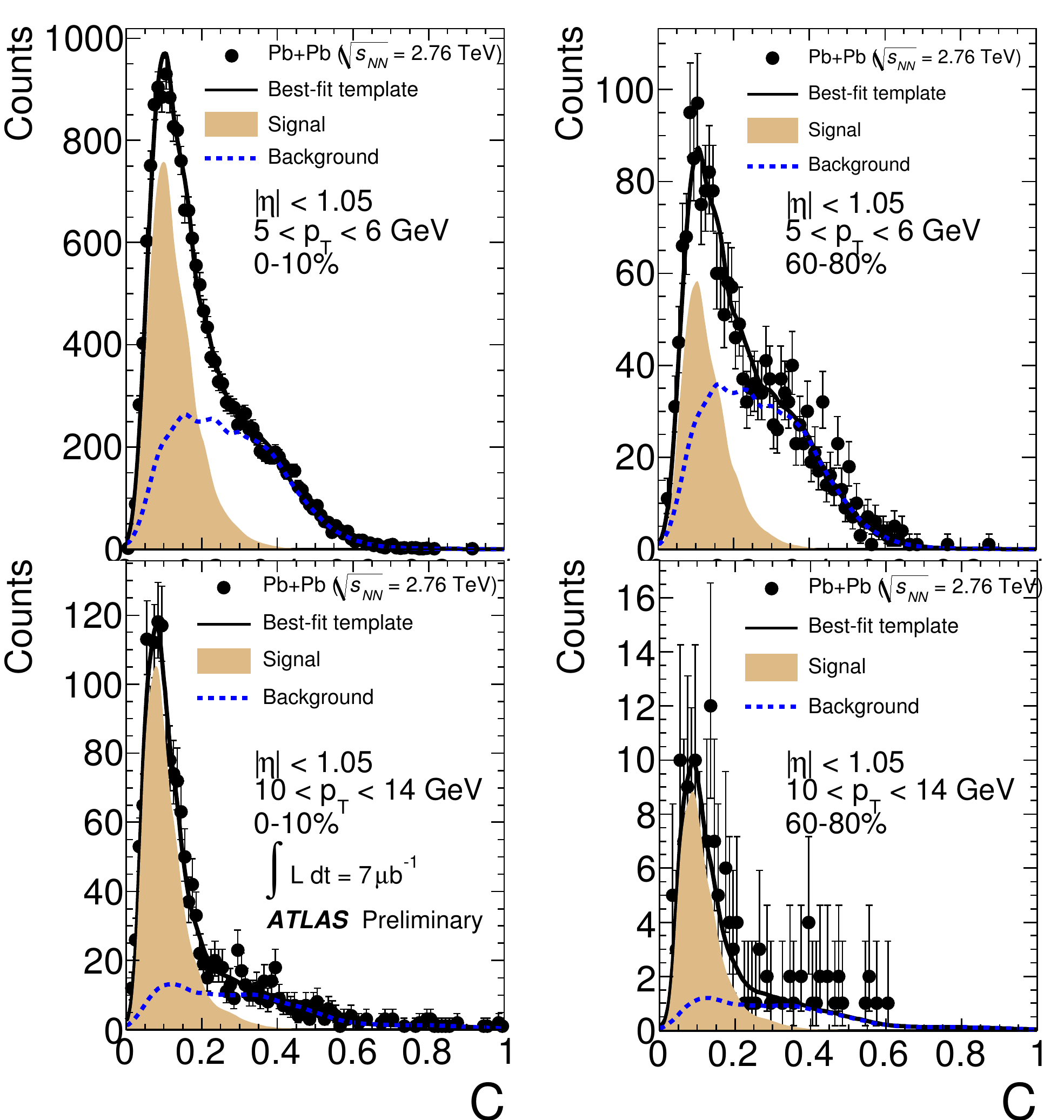} &
    \includegraphics[width=0.45\textwidth]{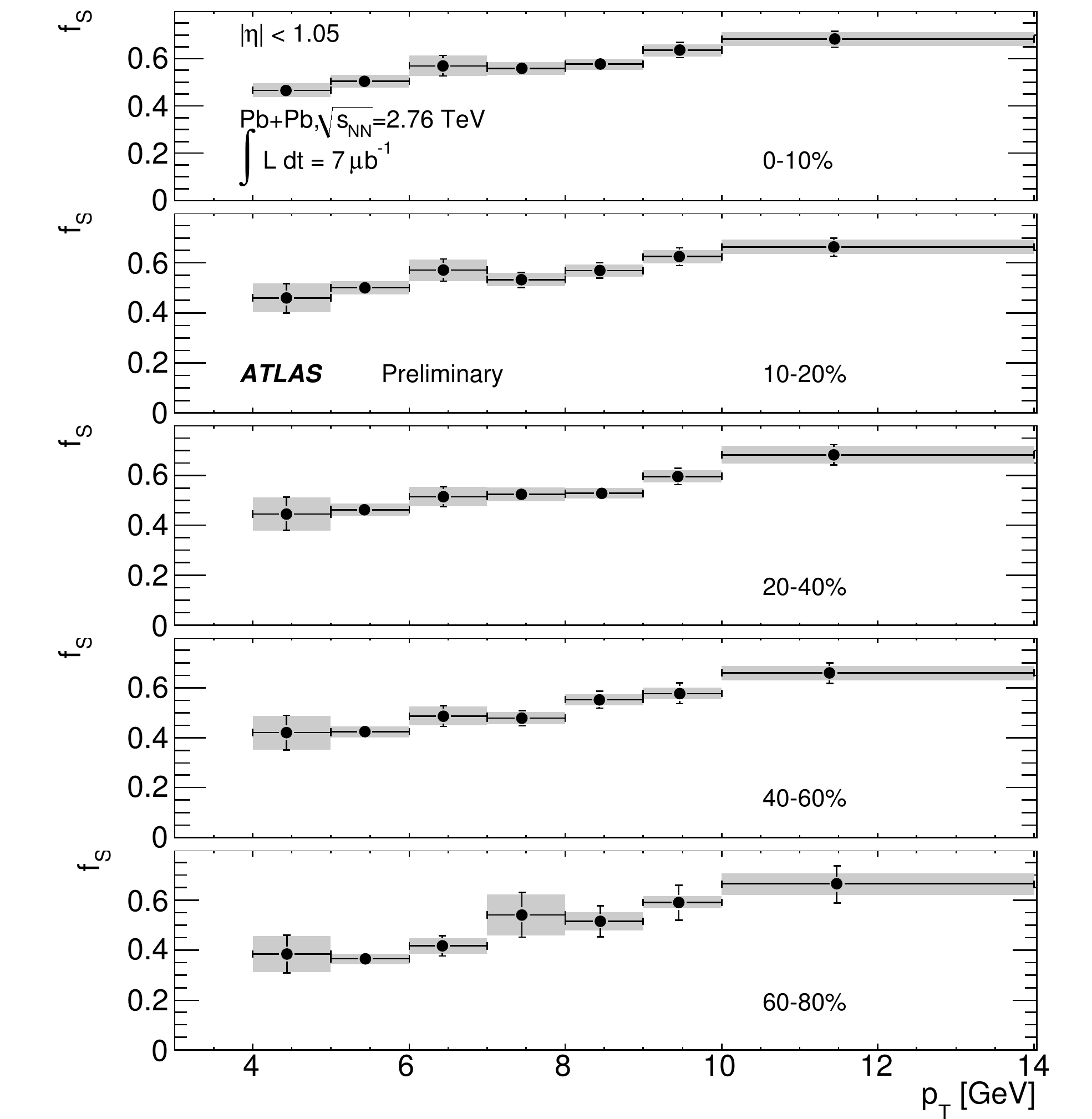} \\
  \end{tabular}
  \caption{
    {\bf Left: }Examples of the template fits to measured
    \Comp\ distributions (see text) in 0-10\% (left) and 60-80\% (right)
    centrality bins for two \pT\ intervals: $5 < \pT < 6$~\GeV\ (top) and
    $10 < \pT < 14$~\GeV\ (bottom). The curves show the contributions of
    signal and background sources based on the corresponding
    \Comp\ distributions obtained from MC including the shift,
    re-scaling, and smearing modifications (see text).
    {\bf Right: }
    Fractions of signal muons, \fs, in the measured muon yields as a function of
    muon \pT\ in different bins of collision centrality. 
    The points are shown at the mean transverse momentum of the muons in
    the given \pt\ bin.
    The shaded boxes indicate systematic errors, 
    which can vary from
    point-to-point. The error bars show combined statistical and
    systematic uncertainties. Please refer to~\cite{ATLAS-CONF-2012-050}
    for details.
  }
  \label{fig:templatefits}
\end{figure}
The signal and background distribution of \Comp\
were obtained separately from the MC and used to
perform template fits to the \Comp\ distribution of the data.
Specifically, a probability density distribution
for \Comp, $\mathrm{d}P/\mathrm{d}\Comp$, is
formed assuming that a fraction, \fs, of the measured muons is signal, 
$
\dfrac{\mathrm{d}P}{\mathrm{d}\Comp} = \left.\fs \dfrac{\mathrm{d}P}{\mathrm{d}\Comp}\right|_{\mathrm{S}} + 
(1 - \fs)\left.\dfrac{\mathrm{d}P}{\mathrm{d}\Comp}\right|_{\mathrm{B}}$,
where $\mathrm{d}P/\mathrm{d}\Comp|_{\mathrm{S}}$ and
$\mathrm{d}P/\mathrm{d}\Comp|_{\mathrm{B}}$ represent the Monte Carlo
\Comp \ distributions from signal and background sources,
respectively. 
To account for possible differences in the
\Comp\ distribution between data and MC due to (e.g.) inaccuracies in
the MC description of multiple scattering, the
fits allow for limited adaptation of the MC templates to the data. In
particular, the fit allows a shift and re-scaling of the
\Comp\ distribution, $\Comp' = a + \avgC + b\left(\Comp -
\avgC\right)$, where $\avgC$ is the mean of the \Comp\ distribution.
In addition, a Gaussian
smearing of the MC \Comp\ distribution was included such that the data
\Comp\ distribution was fit to 
$
\dfrac{\mathrm{d}P'}{\mathrm{d}\Comp'} \equiv \left(\fs \left.\dfrac{\mathrm{d}P}{\mathrm{d}\Comp'}\right|_{\mathrm{S}} + 
(1 - \fs)\left.\dfrac{\mathrm{d}P}{\mathrm{d}\Comp'}\right|_{\mathrm{B}}\right) \otimes
\dfrac{e^{-{\mathcal C}'^2/2\sigma^2}}{\sqrt{2\pi}\sigma}
$.
A kernel estimation method \cite{Cranmer:2000du} was used to produce
the unbinned probability density distribution, $dP'/d\Comp'$ from the
MC signal and background samples. The signal fraction, \fs, was then
evaluated for each centrality and muon \pT\ bin using unbinned maximum
likelihood fits with four free parameters, \fs, $a$, $b$, and
$\sigma$. 
Examples of the
resulting template fits are shown in the left panel of Fig.~\ref{fig:templatefits} for
the 0-10\% and 60-80\% centrality bins and for two muon
\pT\ intervals. Typical fitted values for the $a, b,$ and $\sigma$ parameters 
are, $a \sim 0.02$, $b \sim 0.95-1.05$, and $\sigma \sim 0.002$.
The combination of the MC signal and background
\Comp\ distributions are found to describe well the measured
\Comp\ distributions. 
The description remains good even if the
adaptation of the template is not implemented,
and the resultant variation in the fit
contributes to the systematic error,
as discussed below. %% As can be seen in
%% Fig.~\ref{fig:templatefits}, the  signal and background
%% contributions to the \Comp\ distribution are quite distinct at
%% higher \pT. At lower muon \pT\ values the separation between the two
%% contributions is less clear and the estimation of \fs\ is potentially
%% more susceptible to systematic uncertainties. 

The signal muon fractions extracted from the template fits for all
\pT\ and centrality bins are shown in the right panel of Fig.~\ref{fig:templatefits}. 
The
statistical uncertainties on \fs\ from the fits represent $1\sigma$
confidence intervals that account for the limited statistics in the
data \Comp\ distributions and correlations between the fit parameters. 
The uncertainties in the fit results due to the finite MC statistics were
evaluated using a pseudo-experiment technique in which new pseudo-MC
\Comp\ distributions with the same number of counts as the original MC
\Comp\ distributions were obtained by statistically sampling the MC
distributions. The resulting pseudo-MC distributions were then used to
perform template fits. The procedure was repeated eight times for each
\pT\ and centrality bin and the standard deviation of the resulting
\fs\ values in each bin was combined in quadrature with the
statistical uncertainty from the original fit to produce a combined
statistical uncertainty on \fs. The fractional uncertainties on \fs\ due
to MC statistics are found to be below $2\%$ and are typically much
smaller than the uncertainties from the template fits. 

The sensitivity of the measured \fs\ values to the adaptation of the
MC \Comp\ distributions to the data was evaluated by performing the
fits without implementing the adaption, i.e.\ by fixing 
the $a$ and $\sigma$ parameters to zero and $b$ to one.
To test the sensitivity to the primary hadron composition in the 
MC, and in particular to the relative proportion of kaons and pions (\Ktopi),
which may differ from that in the data, 
new MC background
\Comp\ distributions were obtained by separately doubling the
$\pi$ and $K$ contributions. 
Potential systematic uncertainties resulting from the template fitting
procedure were evaluated using a simple cut-based procedure applied to the
\Comp\ distributions. The total uncertainties from the above considerations
are around 7$\%$ and 5$\%$ for 7 -- 8 GeV and 10 -- 14 GeV bins in the 
0 -- 10 $\%$
centrality interval, and are around 19$\%$ and 6$\%$ for 
7 -- 8 GeV and 10 -- 14 GeV bins in the 60 -- 80 $\%$ centrality interval. 

\begin{figure}
  \begin{tabular}{cc}
    \includegraphics[width=0.45\textwidth]{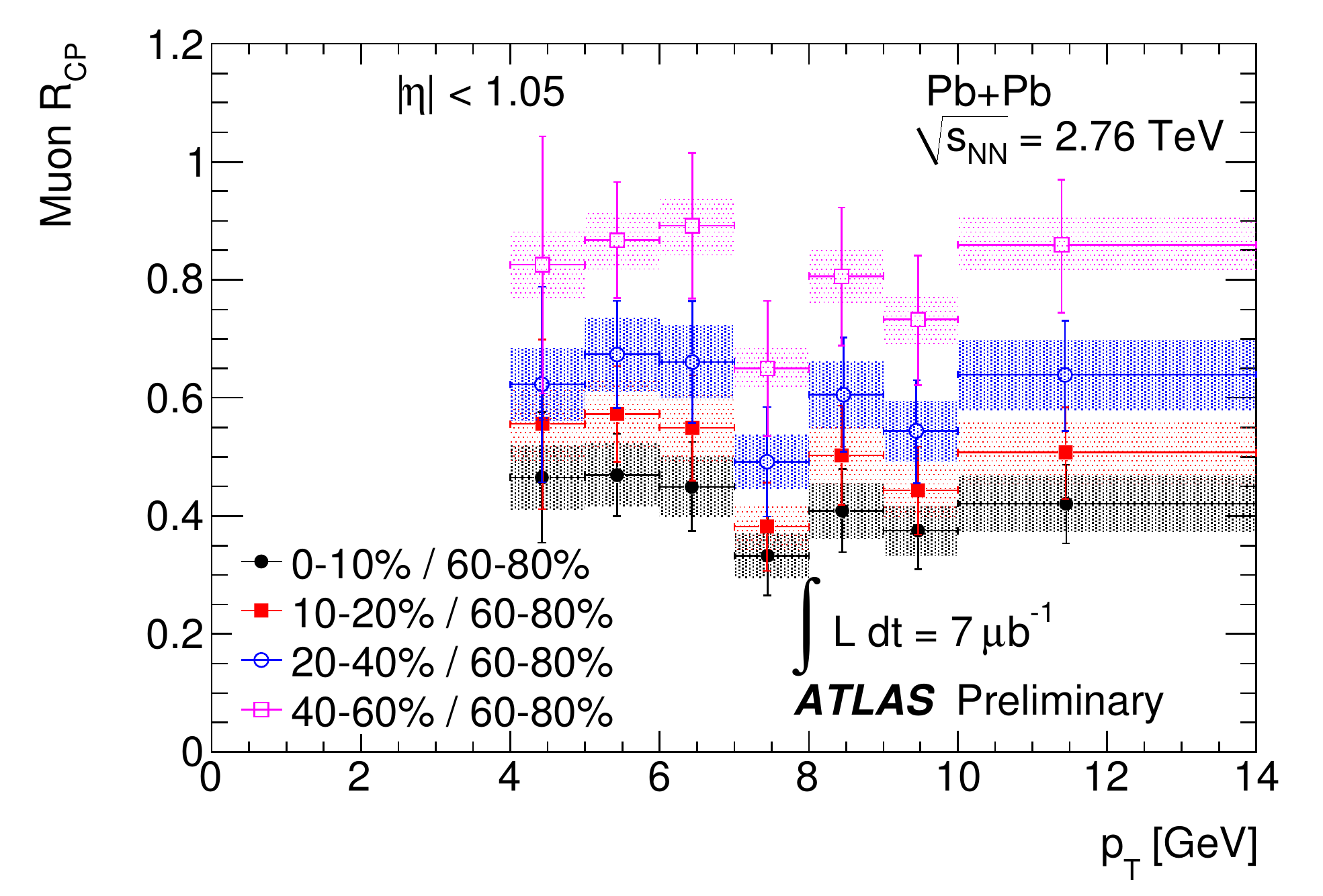} &
    \includegraphics[width=0.45\textwidth]{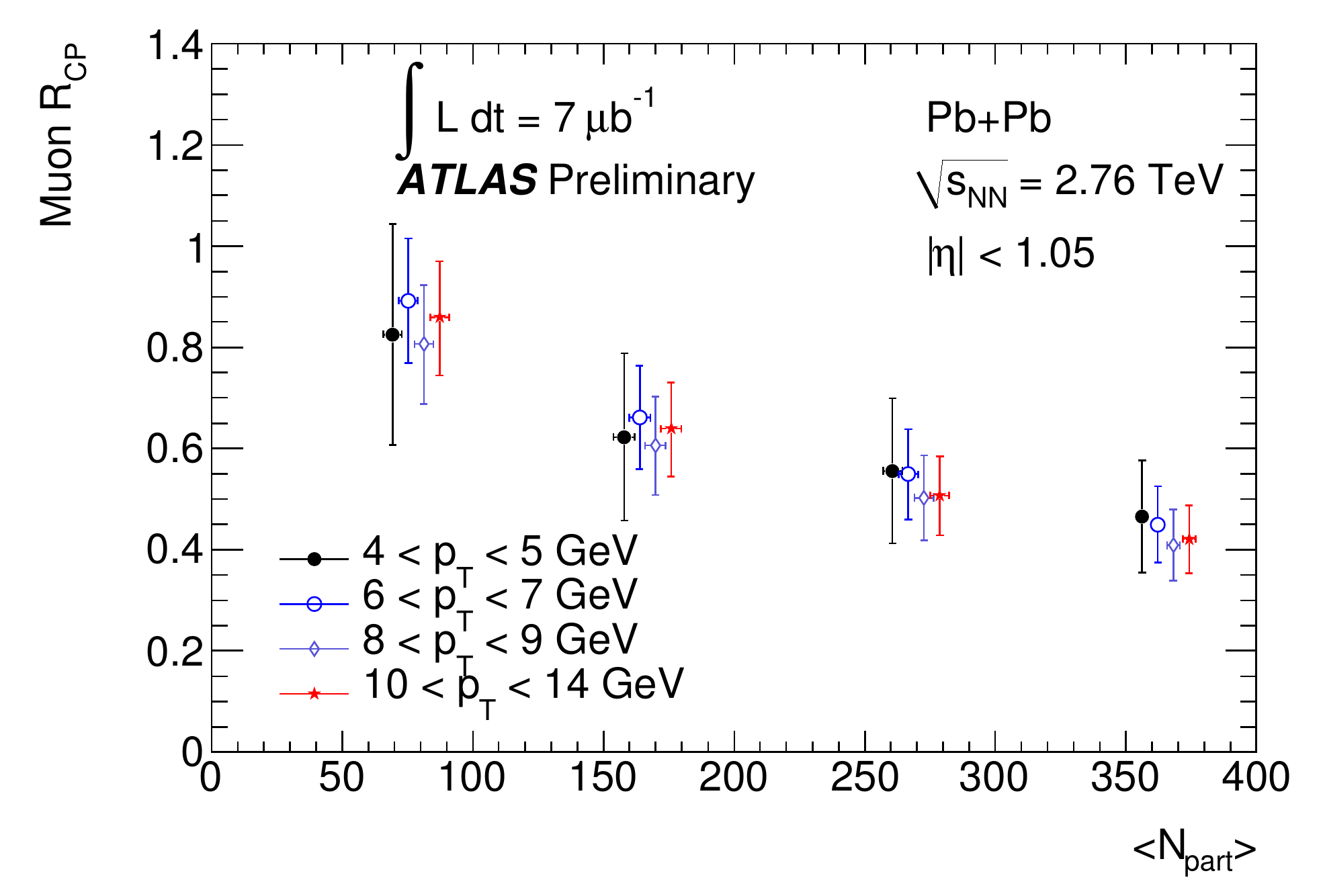}  \\
  \end{tabular}
  \caption{
    {\bf Left: } Muon \Rcp\ as a function of \pT\ for different centrality bins. 
    The points are shown at the mean transverse momentum of the muons in
    the given \pt\ bin. The error bars include
    both statistical and systematic uncertainties. The contribution of the
    systematic uncertainties from \Rcollcent\ and $\varepsilon$, which are
    fully correlated between \pT\ bins, are indicated by the shaded boxes.
    {\bf Right: } Muon \Rcp\ as a function of \ANpart\ for different bins in muon \pT. 
    The error bars show combined statistical and systematic
    uncertainties. The sets of points for the different \pT\ bins are
    successively displaced by $\Delta \Npart = 6$ for clarity of presentation.
    Please refer to~\cite{ATLAS-CONF-2012-050} for details.
  }  
  \label{fig:rcp}
\end{figure}

The resulting central-to-peripheral values, \Rcp , are shown in 
the left panel of Fig.~\ref{fig:rcp}. The
\Rcp\ values vary weakly with \pT , and the points for each
centrality interval are consistent with a \pT-independent \Rcp\ 
within the uncertainties on the points. The \Rcp\ varies strongly with
centrality, increasing from about 0.4 in the 0-10\% centrality bin to
about 0.85 in the 40-60\% bin. 
The variation of \Rcp\ with centrality as
characterized by the average number of participants, \ANpart, is shown in
the right panel of Fig.~\ref{fig:rcp} for the different \pT\ bins included in the
analysis. The \Rcp\ decreases smoothly from peripheral to central
collisions. The centrality dependence is observed to be the same for
all \pT\ bins within the experimental uncertainties.

\section{Conclusions}
This paper has presented  ATLAS measurements of muon production
and suppression in $\sqrtsnn = 2.76$~\TeV\ \PbPb\ collisions in a
transverse momentum range dominated by heavy flavour decays, $4 < \pT <
14$~\GeV, and over the pseudo-rapidity range $|\eta| < 1.05$. The
fraction of prompt muons was estimated using template fits to the
distribution of a quantity capable of distinguishing statistically 
between signal and background. 
The centrality dependence of muon production
was characterized using the central-to-peripheral ratio,
\Rcp, calculated using the 60-80\% centrality bin as a peripheral
reference. The results for \Rcp\ indicate a factor of about 2.5
suppression in the yield of muons in the most central (0-10\%)
collisions compared to the most peripheral collisions included in the analysis. 
No significant variation of \Rcp\ with muon \pT\ is
observed. The \Rcp\ decreases smoothly from peripheral to central
collisions. 
%% The central muon \Rcp\ for $10 < \pT < 14$~\GeV\ agrees with
%% the central (0-10\%) non-prompt \Jpsi\ \RAA\ measured by CMS. 
%%  The central \Rcp\ is 1.5-2 times larger than that
%% measured for charged hadrons in a comparable \pT\ range and using
%% comparable, though not identical, centrality ranges. The
%% central \Rcp\ indicates weaker suppression than observed in the
%% semi-leptonic electron \RAA\ measured over the same \pT\ range in
%% $\sqrts = 200$~\GeV\ \AuAu\ collisions at RHIC.



%% References


\bibliographystyle{elsarticle-num}
\bibliography{Proceeding.bib}

\begin{thebibliography}{10}
\expandafter\ifx\csname url\endcsname\relax
  \def\url#1{\texttt{#1}}\fi
\expandafter\ifx\csname urlprefix\endcsname\relax\def\urlprefix{URL }\fi
\expandafter\ifx\csname href\endcsname\relax
  \def\href#1#2{#2} \def\path#1{#1}\fi

\bibitem{Majumder:2010qh}
A.~Majumder, M.~Van~Leeuwen, Prog.~Part.~Nucl.~Phys. A66 (2011) 41--92.

\bibitem{Wang:1994fx}
X.-N. Wang, M.~Gyulassy, M.~Plumer, Phys.~Rev. D51 (1995) 3436--3446.

\bibitem{Baier:1998yf}
R.~Baier, Y.~L. Dokshitzer, A.~H. Mueller, D.~Schiff, Phys.~Rev. C58 (1998)
  1706--1713.

\bibitem{Gyulassy:2000fs}
M.~Gyulassy, P.~Levai, I.~Vitev, Phys.~Rev.~Lett. 85 (2000) 5535--5538.

\bibitem{Djordjevic:2003zk}
M.~Djordjevic, M.~Gyulassy, Nucl.~Phys. A733 (2004) 265--298.

\bibitem{Wicks:2005gt}
S.~Wicks, W.~Horowitz, M.~Djordjevic, M.~Gyulassy, Nucl.~Phys. A784 (2007)
  426--442.

\bibitem{Dokshitzer:2001zm}
Y.~L. Dokshitzer, D.~Kharzeev, Phys.~Lett. B519 (2001) 199--206.

\bibitem{Adare:2006nq}
A.~Adare, et~al., Phys.~Rev.~Lett. 98 (2007) 172301.

\bibitem{Abelev:2006db}
B.~Abelev, et~al., Phys.~Rev.~Lett. 98 (2007) 192301, erratum-ibid. 106 (2011)
  159902.

\bibitem{Adare:2010de}
A.~Adare, et~al., Phys.~Rev. C84 (2011) 044905.

\bibitem{Djordjevic:2011tm}
M.~Djordjevic, Phys.Rev. C85 (2012) 034904.

\bibitem{Gossiaux:2008jv}
P.~Gossiaux, J.~Aichelin, Phys.~Rev. C78 (2008) 014904.

\bibitem{Uphoff:2011ad}
J.~Uphoff, O.~Fochler, Z.~Xu, C.~Greiner, Phys.~Rev. C84 (2011) 024908.

\bibitem{Horowitz:2008ig}
W.~Horowitz, M.~Gyulassy, J.~Phys.~G G35 (2008) 104152.

\bibitem{Aad:2008zzm}
{ATLAS Collaboration}, JINST 3 (2008) S08003.

\bibitem{ATLAS-CONF-2012-050}
{ATLAS Collaboration}, {ATLAS-CONF-2012-050}.

\bibitem{ATLAS:2011ah}
{ATLAS Collaboration}, Phys.~Lett.~ B707 (2012) 330--348.

\bibitem{atlassim}
{ATLAS Collaboration}, Eur.~Phys.~J. C70 (2010) 823--874.

\bibitem{Agostinelli:2002hh}
S.~Agostinelli, et~al., Nucl.~Instrum.~Meth. A506 (2003) 250--303.

\bibitem{Wang:1991hta}
X.-N. Wang, M.~Gyulassy, Phys.~Rev. D44 (1991) 3501--3516.

\bibitem{Cranmer:2000du}
K.~S. Cranmer, Comput.~Phys.~Commun. 136 (2001) 198--207.

\end{thebibliography}






% \end{thebibliography}

\end{document}